\documentclass[aps,onecolumn,showpacs,prl,12pt]{revtex4}

\usepackage{graphicx}
\usepackage{amsmath}
\usepackage{dcolumn}
\usepackage{epsfig}
\usepackage{natbib}
\usepackage{amsfonts}
\usepackage{amssymb}

\begin{document}

\title{ Dynamics of femtosecond magnetization reversal induced by circularly polarized light in the presence of 
fluctuations and dissipation}

\author{A. Rebei}
\email{arebei@mailaps.org}
\author{J. Hohlfeld}
\email{julius.hohlfeld@seagate.com}
\affiliation{Seagate Research Center, Pittsburgh, PA 15222}

\begin{abstract}
Magnetization reversal by a femtosecond circularly polarized laser pulse 
has been recently demonstrated in rare-earth doped 
transition metals (RE-TM). The switching mechanism has been attributed to an 
inverse Faraday effect and thermal effects. Based on the parameters provided in the 
experimental work, we show that this claim is unlikely 
to give rise to femtosecond reversal. Using a hybrid itinerant-localized 
picture of the RE-TM system, we propose
 a new mechanism that requires the presence of the rare earth 
element to reduce the symmetry of the system  as well as
 a strong   enhancement of
spin-orbit coupling between  the d electrons and the 
f moments in the presence of the laser. Our model does not require
 the heating close to the  Curie temperature of the 
sample. 
\end{abstract}
\date{\today}
\maketitle

Using perpendicular magnetized $Gd_{22}Fe_{74.6}Co_{3.4}$ thin films, it was 
shown that a $40\,$fs circularly polarized laser pulse with a wavelength
of $800 \,$nm is capable of switching the magnetization \cite{rasing}. What 
is remarkable about this important and unexpected result is that it was all-optical 
and no external fields were needed. Instead the switching was critically dependent 
on the chirality and the 
intensity of the laser pulse.  The minimum intensity of the laser needed to accomplish 
reversal was 
of the order of 
$10^{11}$ W/cm$^2$ which is still too low for the direct exchange 
of the angular momentum between the laser and the magnetization
 to be significant.  The authors of Ref.\,\cite{rasing} argued that a magnetic 
field generated by the inverse Faraday effect, $H_{\rm IFE}$, in combination with the enhanced 
susceptibility at elevated temperatures 
is responsible for the switching. They estimated $H_{\rm IFE}$ to be of the order of $10$\,T and the 
switching 
temperature $T_{\rm SW}$ to be close to the Curie point $T_{\rm C}$. However an equally valid 
reinterpretation of their data leads to $T_{\rm SW} \approx (T_{\rm C}-T_{\rm room})/2$ \cite{julius}.
The dynamics of the reversal was not investigated in \cite{rasing}, but it is believed to 
occur on the femtosecond time scale. This belief is based on the fact that controlled switching 
is only possible if it is faster than relevant energy and phase relaxation times.

In this paper, we first discuss the ideas put forward in Ref.\,\cite{rasing} in a more quantitative way and 
show that their picture can not explain  femtosecond magnetization reversal. After that, we suggest 
qualitatively a different  richer picture of  the dynamics in the system and point out 
the need to go beyond LLG to study magnetization dynamics on the 
femtosecond time scale. The necessity stems from the fact, that strong magnetic fields
 generated by the laser are alone 
 not sufficient to drive femtosecond 
magnetization reversal. Instead it is the simultaneous  optical enhancement of dissipation that
 enables ultrafast switching. This dissipation is made possible by the fast relaxation 
of  the hot electrons which requires the full account of the non-adiabatic dynamics 
in the laser-spin-electrons system. This latter aspect of the problem is treated in detail in this paper since 
it is of tremendous importance to any non-equilibrium system and in particular fast spin dynamics
in systems 
driven far from equilibrium by a laser pulse.
Consequently,
 a self-consistent treatment of dissipation and fluctuations is required which is also included here. 
Our  method of solution goes beyond the Born approximation and is non-perturbative which is needed to 
accurately treat spin dynamics
 in an interacting environment as discussed by DiVincenzo and Loss \cite{vincenzo}. It is based on 
the two-point effective action \cite{rebei2} and 
provides a systematic way to obtain solutions more accurate than those of the  
Bloch equations \cite{bloch}.

Before we discuss our model and method of solution, we first show that the ideas proposed in Ref.
\cite{rasing}, which are based on LLG, can not form a basis for the understanding of the fast switching 
of the magnetization. The LLG equation is a phenomenological equation written for magnetization close to 
equilibrium and accounts for damping in the simplest possible way
\begin{equation}
\frac{d{\bf M}}{dt}=- \gamma {\bf M}\times  {\bf H}^{eff}
+\alpha  {\bf M}\times  {\frac{d \bf M}{d t} } .\label{ll}
\end{equation}
The internal field and the external laser field, ${\bf {H}} _l=H_l (\cos \omega_l t, \sin \omega_l t, 0 )$, are part of ${\bf H}^{eff}$. In a frame rotating 
with the frequency $\omega_l = 10^{15} Hz$, $\bf{H} _l$ becomes time independent and the lab frame time derivative $\frac{d{\bf M}}{dt}$ 
becomes $\frac{d{\bf M}}{dt}- \omega_l \bf{z} \times \bf{M}$. Hence in the rotating frame, the effective field is ${\bf H}^{eff} + {\bf H}_B$, where  ${\bf H} _B =  -\omega_l/\gamma {\bf z}$ is the equivalent of the Barnett field which is of the order of $5\times 10^7$ Oe. This shows that the magnetization in the presence 
of the laser precesses at almost five orders of magnitude faster than in typical FMR experiments.  
The LLG equation has been successful in reproducing most magnetization data with 'reasonable' damping constants $\alpha$ but it does a poor job in systematically accounting for damping in different correlated data sets. An example of such data is the measurement  of damping in rare-earth doped transition metals (RE-TM) as a function of the orbital momentum of the impurities \cite{bailey}.  

\begin{figure}[th]
\mbox{\epsfig{file=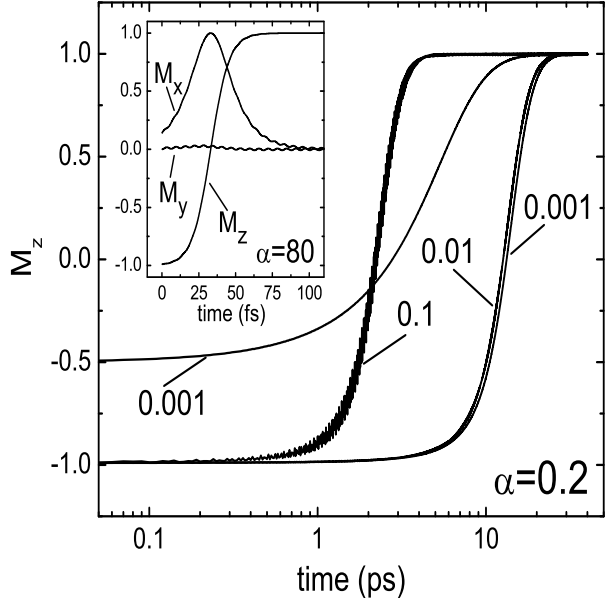,height=8 cm,width=8 cm}}
\caption{ Dynamics of the perpendicular component $M_z$  
in an inverse Faraday field $H_{IFE}=10$ T. The curves are
plotted for various laser fields   $H_l/H_B=0.1$,  $0.01$, and $0.001$ with initial $M_z(t=0)=-1$. The
curve strating at  $M_z(0)=-0.5$ emulates the effect of elevated temperatures on the reversal for  $H_l/H_B=0.001$. The LLG damping is $\alpha=0.2$. The inset is for  $H_l/H_B=0.001$ and the unphysical damping $\alpha=80.0$. }
\label{llg}
\end{figure}

The dynamics of the magnetization reversal as obtained from
 Eq.\,\ref{ll} with $\alpha = 0.2$ and $H_{\rm IFE} = 10$ T
(experimental values of Ref. \cite{rasing}) is shown in Fig.\,\ref{llg} for various laser fields.  The 
laser intensity $H_l/H_B=0.001$  corresponds to the experimental power and  
shows slow reversal within  $\approx 20\,$ps. 
The faster reversal curves for $H_l/H_B=0.01$ and $0.1$ is 
  due  to the Zeeman  interaction with the laser but requires intensities well above the damage threshold.
To achieve  femtosecond switching at reasonable laser powers, the damping would have to be 
increased to 
 non-physical values of  $\alpha = 80.0$ (cf.\,inset of figure \ref{llg}). This calls for a different way to 
include dissipation in fast spin dynamics.  Heating GdFeCo to temperatures
some tens of degrees below the Curie point also 
does not help achieve faster than picosecond reversal. The role of temperature is to increase 
 the fluctuations in the initial values of $M_z$ which will help speed up the reversal but not to the 
extent of being fast at the femtosecond scale. The 
Curve with initial magnetization $M_z ( 0 )=-0.5$ shows the effect of 
increasing temperature on the switching speed.

In the rest of this paper, we give a more careful treatment of the two aspects of time dependent fields and of damping in a simplified model accompanied by a qualitative discussion of the various factors involved in the reversal.

\begin{figure}[th]
\mbox{\epsfig{file=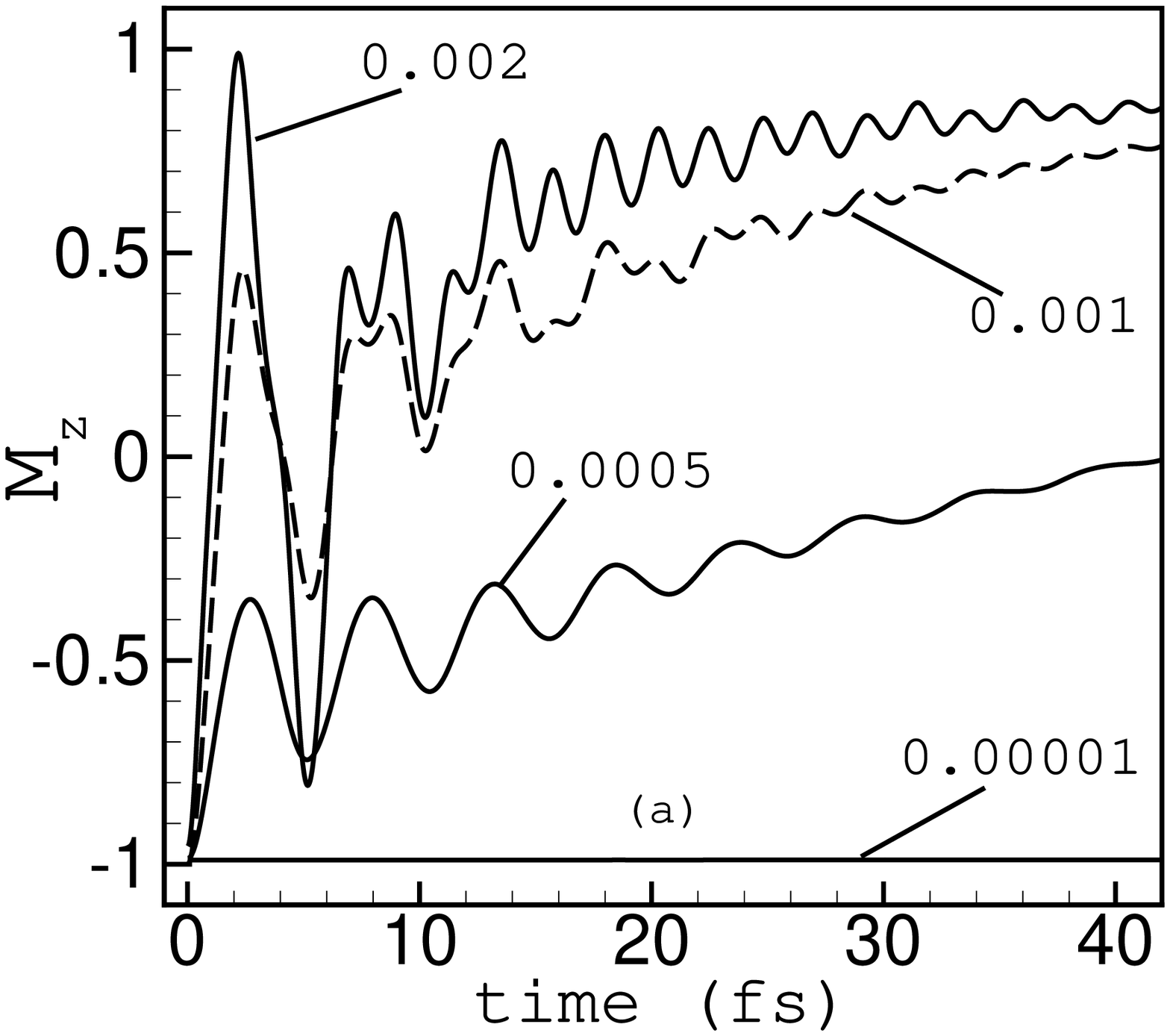,height=6 cm}}
\mbox{\epsfig{file=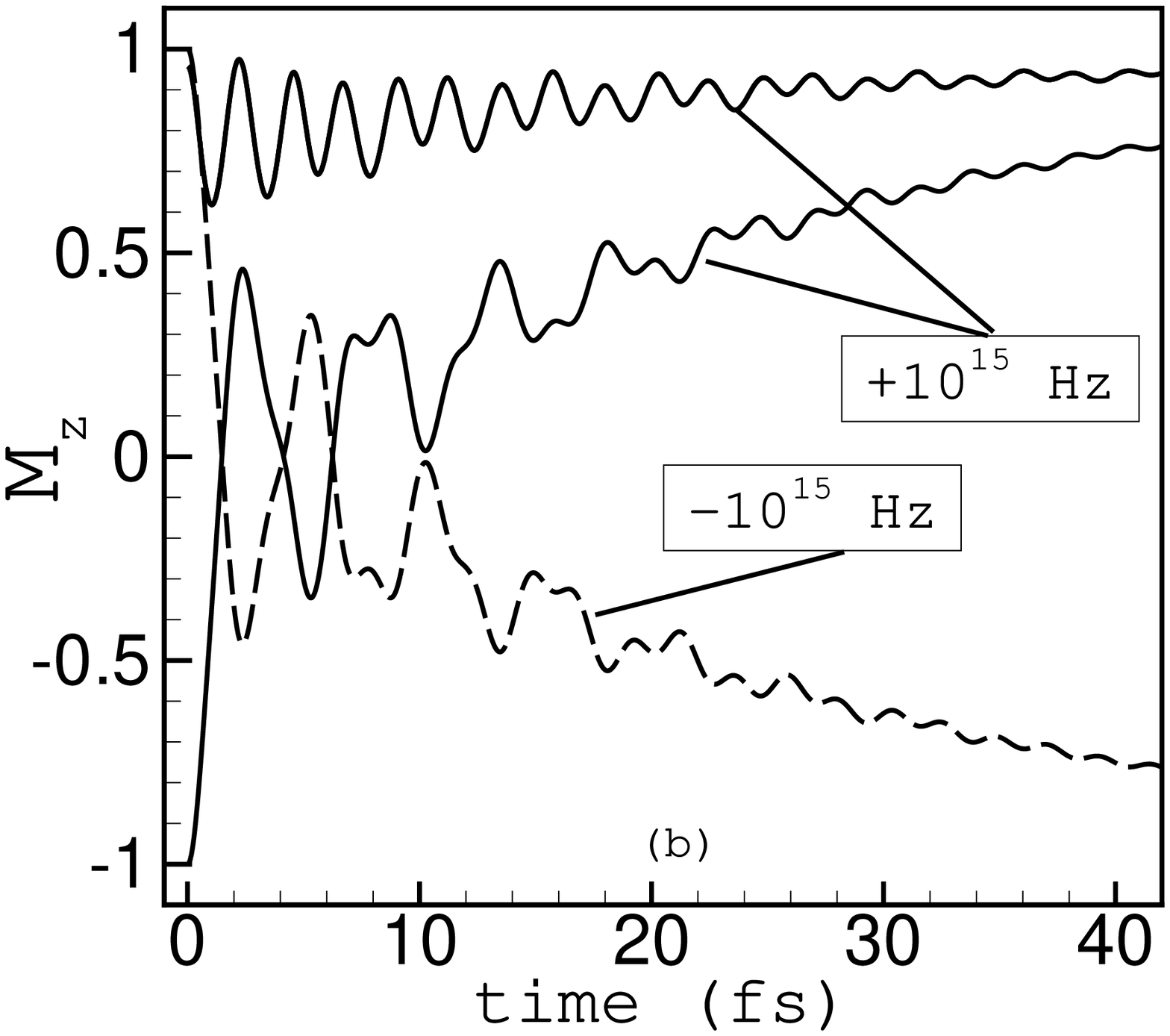,height=6 cm}}
\caption{ (a): Dynamics of  the perpendicular  component of the magnetization 
${M}_z$  induced 
by circularly polarized light $(E_l,H_l)$ with positive frequency $\omega_l=10^{15}$\,Hz and various intensities
$H_l/H_B=0.00001$, $0.0005$, $0.001$, and $0.002$. The initial magnetization is $M_z(0)=-1$. (b): The dependence of switching on the chirality and initial conditions $M_z= \pm 1$ for $H_l/H_B=0.001$.  All results are for
 $\omega_0 = 0.8 \;\omega_l$, $\Gamma=0.2$, and 
$A=10^4 \,$Oe. $H_l=10^{-3} H_B$ corresponds to a power of $\approx 5 \cdot 10^{11}$W/cm$^2$.}
\label{a1}
\end{figure}

The excitation of GdFeCo by the laser  unquenches the orbital angular momentum of the hot electrons which are mainly d electrons provided by Fe. But the d electrons also carry magnetization and hence  a strong interplay of orbital magnetization and spin magnetization in this system is expected. The Hamiltonian is
\begin{eqnarray}
\mathcal{H} & = & \mathcal{H}_d + \mathcal{H}_{ph} + \mathcal{H}_i \, , \label{ham}\\
\mathcal{H}_i & = & -e \mathbf{E}(t)\cdot \mathbf{r} -\mathbf{H}(t)\cdot \left( \mu_d \mathbf{s}  + \mu_f \mathbf{S}_f \right)   \\
& & + J \mathbf{s} \cdot \mathbf{S}_f + \lambda \mathbf{l} \cdot \mathbf{s} + \lambda_{so} \mathbf{l} \cdot \mathbf{S}_f \; . \nonumber
\end{eqnarray}
The Hamiltonian $\mathcal{H}_d$ is that of the d (and s, p) electrons of both Fe and Gd, 
$\mathcal{H}_{ph}$ is the Hamiltonian of the phonons and $\mathcal{H}_i$ is the interaction 
Hamiltonian which includes the dipole interaction, the Zeeman term with both d and f electrons, 
the spin-orbit coupling term of the d electrons with coupling constant $\lambda$, the 
antiferromagnetic coupling between the d and f electrons and finally the effective spin-orbit 
coupling between the orbit of 
the itinerant d electrons and the local f moment of Gd that gives rise to the perpendicular 
anisotropy in GdFeCo. This latter term has many common features to the well known 
Rashba coupling which gives rise to dissipationless spin currents in time-dependent 
or inhomogeneous fields \cite{rebeiolle}. It is the smallest of interactions in the absence 
of the laser but it lowers the symmetry of the effective field seen by the d electrons and 
stores the memory of the chirality of the light after the laser is turned off. The enormous 
complexity of the reversal process in GdFeCo is only partly reflected in this Hamiltonian 
and since no sufficient real-time data are available to warrant detailed calculations based 
on this model, we present here a calculation based on a further idealized Hamiltonian that 
still has the necessary elements needed for a femtosecond reversal process. Our reversal 
process is based on ideas similar to the (mechanical) Barnett effect \cite{barnett} which says 
that an external torque applied to an iron rod affects its magnetization by acquiring a component 
along the axis of the torque.  The torque in our case will be provided by the electric field of the 
laser as it acts on the hot d electrons. However the magnetic field of the laser also provides a 
similar Barnett contribution through the time-dependent Zeeman interaction with the d and 
f moments. The phonons through the spin-orbit coupling and the dipole terms provide the 
dissipation required for the reversal process. Polarization in the presence of rotating fields 
and dissipation has long been recognized in gravitational physics \cite{hawking},  accelerator 
physics \cite{bell} and in semiconductors \cite{edelstein}. Therefore there is no surprise that 
our model will give rise to polarization induced by a circularly polarized laser. In order to 
give a clear discussion of the optical Barnett-like effect, 
we introduce the effective Hamiltonian, $\mathcal{H}^{\rm eff}$,  for 
 the spin degrees of freedom deduced from
Eq.\,\ref{ham} and treat the orbital degrees of 
freedom and the effect of $\mathbf{E}_l$ on them 
as part of the environment. $\mathcal{H}^{eff}$  
 includes energy exchange between three different sub-systems
(cf.\,Fig.\,\ref{a1}, inset) and is $(\hbar = 1)$    
\begin{equation}
\mathcal{H}^{eff}=\mathcal{H}_s+\mathcal{H}_q+ \mathcal{H}_{sql}+
\mathcal{H}_Q
\end{equation}
The spin Hamiltonian, $\mathcal{H}_s= -\gamma \mathbf{H}_l \cdot \mathbf{S}  - \frac{1}{2}A S
_{z}^{2}$, includes the interaction with 
 $\mathbf{H}_l$ and an axial anisotropy 
term. The $\mathbf{S}$'s represent the effective spin degrees of freedom in the 
GdFeCo 
system. The anisotropy, which is
most relevant after the laser is turned off, 
is taken in the mean field approximation. The gyromagnetic ratio $\gamma$ is 
assumed positive. The 
Hamiltonian, $\mathcal{H}_q=\frac{\mathbf{p}^2}{2} 
+ \frac{\omega_0^2}{2} \mathbf{q}^2 $, represents 
 a single optical electronic mode 
 with frequency $\omega_0$ close to the laser frequency $\omega_l$ \cite{feynman}. The
Hamiltonian 
$\mathcal{H}_Q$ is that of the 
macroscopic bath as defined in \,\cite{schwinger1}. The interaction Hamiltonian 
$\mathcal{H}_{sql}= - \lambda
\mathbf{S}\cdot \mathbf{q}  -\mathbf{q}\cdot \mathbf{Q}$, includes an effective  
   linear coupling of the spin to the optical mode through spin-orbit coupling, which is 
taken isotropic for simplicity, and a linear 
coupling of the optical mode to the phonons $\bf Q$. The material dependent parameter
 $\lambda=\lambda_0 
+ \lambda_1 E_l^2/E_c^2$ is an effective spin-orbit contribution with $\lambda_0 \approx 0.001$\, eV, 
$\lambda_ 1 \approx 0.1$\, eV and $E_c$ is the critical field for switching which is taken 
from experiment. In the presence of the laser, an electron acquires additional orbital angular momentum $e^2 E_l^2/m \omega_l^3$ and hence the power dependence of $\lambda$. As
first pointed out in Ref.\,\cite{hubner},  a
 strong spin-orbit coupling in NiO is needed to
 provide a fast relaxation channel for the spin
of the electrons in the femtosecond regime. 
%\begin{figure}[th]
%\mbox{\epsfig{file=energy,height=5 cm,width=5 cm}}
%\caption{The different energy transfer channels involved 
%in the excitation of
% the magnetization $\mathbf{M}$ by the laser $\omega_l$ and its 
% fast relaxation by the macroscopic bath $\mathbf{Q}$ via an optical
% mode $\mathbf{q}$ with energy $\omega_0 \approx \omega_l$.}
%\end{figure}

This effective Hamiltonian is the simplest model that describes well the femtosecond reversal 
and reproduces LLG close to equilibrium 
if $\omega_0$ is much larger than the frequency of the spin \cite{rebeiS}. In the present 
case, $\omega_0$ is very close to that of the laser-driven spin and consequently one of the main 
assumptions underlying LLG, the separation of energy scales between the spin and 
the environment, is broken 
\cite{brown}. Hence we expect non-adiabatic behavior to be important in this system. We will 
show that a careful treatment of the dynamics give non-exponential switching different than the 
one expected from the Bloch equations \cite{bloch} or the LLG equation.  This is the physics 
we would like to address in our analysis of the switching since similar conditions are also present 
in other important
problems such as those in quantum computation  where it may be necessary to go 
beyond the Born approximation to study coherence \cite{vincenzo}.

The coupled 
spin-laser-bath system is better studied in a frame 
$\left(\mathbf{x}_1, 
\mathbf{x}_2,\mathbf{x}_3=\mathbf{z}\right)$ rotating
around the 
$z-$axis 
with  frequency  
 $\omega_l$.  In this frame,  the 
Larmor torque is time-independent and   
the spin Hamiltonian  becomes 
\begin{equation}
\mathcal{H}^{\prime}=-\frac{\gamma H_l}{2}M_{1}  -\lambda \Sigma_i
 M_{i}  b_{i}\left(  t\right) - \frac{\gamma}{2}\left( B_0
 +  H_B \right)M_3
\end{equation}
where   
$B_0= A\left\langle S_{z}\right\rangle $. The
 equations of motion for $\langle \mathbf{M} \rangle$, the
average of the 
spin operator, are 
\begin{eqnarray}
\langle \overset{\cdot}{M}_1 \rangle  & =& -\gamma (B_0 +H_B)\langle 
M_2 \rangle + 2\lambda(\langle M_2 b_{3} \rangle - \langle
M_3 b_{2} \rangle ) \label{eq1} \\
\langle \overset{\cdot}{M}_2 \rangle   & =& \gamma (B_0 +H_B) \langle M_1
 \rangle
 - 
\gamma H_l \langle M_3 \rangle + 2 \lambda ( \langle M_3 b_{1} \rangle -
 \langle M_1 b_{3} \rangle ) \nonumber  \\
\langle \overset{\cdot}{M}_3 \rangle  & =& \gamma H_l \langle  M_2 \rangle - 2 
\lambda \left( \langle M_2 b_{1} \rangle -
\langle M_1 b_{2} \rangle \right), \nonumber
\end{eqnarray}
where $b_i$'s are the harmonic mode variables in 
the rotating frame. The equations of motion for $\mathbf{q}$ are
\begin{equation}
(\frac{d^2}{d t^2}+\omega_0^2)\langle q_i \rangle
 = \langle Q_i \rangle + \lambda \langle \sigma_i \rangle \, , \label{eq2}
\end{equation}
The bath $\mathbf{Q}$ is 
assumed Ohmic and is the source of dissipation in the mode $\mathbf{q}$ 
which we take to be $\Gamma=0.2$. In the adiabatic limit and in the absence of 
anisotropy, the 
system spin plus bath, $\mathcal{H}_s+
\mathcal{H}_Q $, has been treated earlier, see e.g.\,Ref. \cite{smirnov}. Here we 
 study the more complex and experimentally relevant 
case of non-adiabatic magnetization reversal where the inclusion of 
memory effects is necessary. 
  
In 
the rotating
 frame, the instantaneous torque is modified by the Barnett 
field. For visible light with  
$\omega_l \approx 10^{15} Hz$, the optical Barnett field is 
as large as $\approx 10^7 \,$Oe. This 
is larger than most exchange 
fields  
and hence a single spin picture should be adequate  even 
for
the treatment of the interaction of a laser with 
a ferromagnet. Despite its tremendous magnitude, we find that this 
effective  field does not induce femtosecond magnetization 
reversal unless three key requirements are met. First, 
the Barnett field has to be much larger than the laser field 
$\mathbf{H}_l$. Secondly, the magnetization has to be coupled through the 
electric field $\mathbf{E}_l$ 
to at least one optical mode $\mathbf{q}$ of energy $\omega_0 \approx 
\gamma H_B$. Third, the damping of this mode due to its 
interaction with the macroscopic bath $\mathbf{Q}$ has to 
enable efficient energy transfer from the spins to the 
mode $\mathbf{q}$. All three requirements can be met by various 
combinations of the parameters, $H_B, \, H_l , E_l, \, \omega_0$ 
and $\lambda$, but the range of each individual 
parameter depends on the particular choice of the 
others.

 To solve equations \ref{eq1} and \ref{eq2}, we need to calculate the average 
of the product of two operators $\langle \sigma_i(t) q_j(t^\prime)
 \rangle$. For this we need the density matrix of the whole 
system or we may use  the more useful functional formalism instead  
(for a detailed discussion of this method and its application to  
 sd exchange in metals see 
\cite{rebei1} and references therein). The 
generating functional is
\begin{eqnarray}
Z\left[ {\bf J}_{1},{\bf J}_{2}\right]  &=&\left\langle \int D
{\bf \eta }D%
{\bf p}D{\bf q}  \exp \left(   -i\int dt\left( \mathcal{H}_s 
\left( {\bf {\eta},t }\right)
\right.    \right.                    \right.\\ 
& & \left.  \left. \left.
+\mathcal{H}_q \left( {\bf p},{\bf q}\right) +\mathcal{H}_{sql}  -{\bf J}%
_{1}\cdot {\bf q}-{\bf J}_{2}\cdot {\bf \sigma }\right) \right) 
\right\rangle _{Q}
\nonumber
\end{eqnarray}
where the spin variables are for simplicity assumed $1/2$ and are expressed in terms of 
Grassmann variables,
${\bf \sigma }=-\frac{i}{2}\eta \times \eta $  \cite{coleman}. This allows the use 
of Wick's theorem in the path integral 
expansion and $\mathbf{J}_i$ are 
two virtual external sources. The problem then becomes very similar to that of a Fermi gas which 
was discussed with the aid of two point functions \cite{rebei2}. The average values 
are calculated in the same way as in Ref. \cite{rebei2}
\begin{equation}
\left. \frac{\delta ^{2}\ln Z}{\delta {\bf J}_{1}^{i}\delta {\bf J}_{2}^{j}}%
\right| _{{\bf J}_{1}={\bf J}_2=0}=-\left\langle q^{i}\sigma ^{j}\right\rangle
+\left\langle q^{i}\right\rangle \left\langle \sigma ^{j}\right\rangle .
\end{equation}
Before switching on the laser, the phonons and the magnetization are
in equilibrium and are weakly coupled
 since  the electrons are in their ground state with a quenched
 orbital angular momentum. After that, the spin is driven out 
of equilibrium by the laser and $\mathbf{q}$ is treated 
as a perturbation which responds linearly to any changes in the spin. As in the $sd$ exchange 
problem \cite{rebei1}, the 
equations of motion are non-local 
in time and integrating out the optical mode $\mathbf{q}$ gives 
rise to dissipation and fluctuations even at zero temperature.

Using a Runge-Kutta scheme, Eqs.\,\ref{eq1} and \ref{eq2} are 
solved self-consistently. In line with the discussion above, we find that 
switching occurs only for a limited range of mutually dependent parameters.  The 
results in fig.\,\ref{a1} focus on the power dependence of the magnetic response for 
$\omega_l=10^{15}\,$Hz, $\omega_0=0.8 \omega_l$, $\Gamma=0.2$, and $A=10^4\,$Oe. While
the response to laser fields $H_l /H_B \le 10^{-5}$ is negligible on a femtosecond time scale, it 
becomes significant for $H_l/H_B = 5\cdot 10^{-4}$ and ultrafast reversal is found at higher fields 
of $10^{-3}H_B$ and $2\cdot 10^{-3}H_B$. At the highest fields, 
$M_z$ shows an almost instantaneous reversal accompanied by strong oscillations. The 
period of these oscillations is governed mostly by $H_B$ while their decay depends 
on $\Gamma$. The reversal slows down with time and the approach to equilibrium depends 
on the coupling $\lambda$ of the spins to the mode $\mathbf{q}$ which is proportional 
to the power. The curves in the inset show the strong dependence of the reversal on the chirality 
of the laser. It is also important to realize that the simple exponential behavior expected from the Bloch 
equation or the LLG equation for the decay of the z-component is no longer true in our model. Fig. \ref{a1} shows that exponential behavior is only recovered after the laser has been on for several periods. The initial 
response of 
the spin system is therefore highly non-adiabatic  due to the simultaneous action of the laser 
and of the dissipation and fluctuations, which are also enhanced by the laser,  
due to the coupling of the hot electrons to the magnetization.

In summary, we have shown that circularly polarized light
can induce femtosecond magnetization reversal,   
when optical electron-phonon modes with frequencies 
comparable to those of the light are made available by the electric field of the laser through 
spin-orbit coupling. We predict a fast non-exponential reversal that  cannot
be recovered 
 from the modified Bloch equations or the  LLG  equations and should be observable in future
time-resolved experiments. The switching does not
require high temperatures and is insentitive to 
the anisotropy which makes the 
proposed mechanism very attractive to high density magnetic recording. 

\vskip 0.5cm

We thank D. Stanciu, A. Kirilyuk and Th. Rasing for discussing their results with us.

\end{document}